\documentclass[fleqn,twoside]{article}
\usepackage{espcrc2}

\usepackage{graphicx}
\usepackage[figuresright]{rotating}

\def\Journal#1#2#3#4{{#1} {\bf #2}, #3 (#4)}

\def\NPB{{\em Nucl. Phys.} B}
\def\NPBP{{\em Nucl. Phys.} B {\em Proc. Suppl.}}
\def\NPA{{\em Nucl. Phys.} A}
\def\PLB{{\em Phys. Lett.}  B}

\def\PRD{{\em Phys. Rev.} D}
\def\ZPC{{\em Z. Phys.} C}
\def\JHEP{{\em J. High Energy Phys.}}
\def\EPJC{{\em Eur. Phys. J.} C}

\def\epp{\varepsilon^{\prime}}

\def\be{\begin{equation}}
\def\ee{\end{equation}}
\def\bea{\begin{eqnarray}}
\def\eea{\end{eqnarray}}

\newcommand{\AmS}{{\protect\the\textfont2
  A\kern-.1667em\lower.5ex\hbox{M}\kern-.125emS}}

\title{Matching the Electroweak Penguins $Q_7$ and $Q_8$}
\author{Elvira G\'amiz \address[UGR]
{Centro Andaluz de F\'{\i}sica de las 
Part\'{\i}culas Elementales (CAFPE) and \\
Departamento de F\'{\i}sica Te\'orica y
del Cosmos, Universidad de Granada \\Campus de Fuente Nueva, 
E-18002 Granada, Spain}, Joaquim Prades \addressmark[UGR] 
and Johan Bijnens
\address{Department of Theoretical Physics 2, Lund University\\
S\"olvegatan 14A, S 22362 Lund, Sweden}}
       
\begin{document}

\begin{abstract}
\noindent We report on recent advances on the computation
of the matrix elements of the electroweak penguins $Q_7$ and $Q_8$
which are relevant for the $\Delta I=3/2$ contribution to
 $\epp_K$ in the chiral limit.
The matching of scale and scheme dependences between Wilson coefficients
and these matrix elements is done analytically at NLO in $\alpha_S$.
\vspace{1pc}
\end{abstract}

\maketitle
\noindent 
Direct CP violation in $K\to \pi \pi$ is usually parameterized
with
\be
\epp_K \simeq  \frac{i}{\sqrt 2} 
\frac{\rm{Re}(a_2)}{\rm{Re}(a_0)} \,
\left[ - \frac{\rm{Im}(a_0)}{\rm{Re}(a_0)} + 
\frac{\rm{Im}(a_2)}{\rm{Re}(a_2)}\right]  
e^{i \left[\delta_2-\delta_0\right]} ,
\ee
where $ i A_I\equiv {\rm a}_I \exp(i\delta_I)$ 
are $K\to\pi\pi$ isospin invariant amplitudes
and $\delta_I$ are final state interaction (FSI) phases.
 The ratio $\rm{Re}(a_0)/\rm{Re}(a_2)\simeq 22.2$ is an 
experimentally well known quantity.

In the limit $m_u=m_d$ and $\alpha_{QED}^2=0$, and neglecting
the tiny electroweak corrections to Re(a$_2$), one gets
\be
\frac{\rm{Im}(a_2)}{\rm{Re}(a_2)} =
 - \frac{3}{5} \, \frac{F_0^2}{m_K^2-m_\pi^2} 
\, \frac{\rm{Im}(e^2 G_E)}{G_{27}} 
\ee 
including FSI to {\em all} orders in CHPT and 
up to $O(p^4)$ non-FSI corrections \cite{BPP98,CG00,PPS01}. 
The coupling $G_{27}$
modulates the 27-plet operator describing $K\to \pi\pi$
at $O(p^2)$ in CHPT.  
Its value can be obtained from a fit
of $K \to \pi \pi$ and $K\to \pi\pi\pi$ 
to $O(p^4)$ amplitudes \cite{K3pifit}.  

The coupling $G_E$ appears in CHPT to $O(e^2 p^0)$
\cite{BW84}
\be
{\cal L}_{\Delta S=1} = C \, F_0^6 e^2 G_E \, \left(U^\dagger
{\cal Q} U \right)_{23} + O(p^2) (G_{27}, \cdots)  .
\ee
The constant $C$ was defined in \cite{BPP98}, $F_0$ is the pion
decay constant in the chiral limit, ${\cal Q}$
 is a 3 $\times$ 3 matrix collecting
the three light quarks electric charges and $U=\exp( i \sqrt 2 \Phi/F_0)$
with $\Phi$ a 3 $\times$ 3 matrix 
collecting the octet pion, kaon, and eta pseudo-scalar boson fields.

In the Standard Model, there are just two operators contributing
to $\rm{Im} (e^2 G_E)$; namely, the so-called electroweak penguins,
$Q_7$ and $Q_8$. In the chiral limit, 
these operators form a closed system under QCD corrections.
Its anomalous dimensions mixing matrix is known to NLO in  the
NDR and HV schemes \cite{Munich,Roma1}.

There has been recently a lot of work devoted to calculate
$\rm{Im} (e^2 G_E)$, both analytically 
\cite{DG00,NAR01,KPdR01,CDGM01,Atalks} and using lattice
QCD \cite{Roma2,CPPACS,RBC,staggered}.
Here, we report on the work presented in \cite{BGP01}
and to which we refer for explicit formulas and further references.
For a review of the method we used and the matching procedure
see \cite{BPX}. 
Using this method,   one can write exact results for the coupling 
$\rm{Im} (e^2 G_E)$ in terms of integrals 
of full two-point functions in the chiral limit.
They can be related to spectral functions via dispersion relations
and resummation of the effect of all higher dimensional operators
in the OPE of the relevant two-point functions.

The coupling $\rm{Im} (e^2 G_E)$  can be written  as
\bea
\label{result1}
-\frac{3}{5} F_0^6 \, {\rm Im} (e^2 G_E) = 
-  6 \, {\rm Im} C_7(\mu_R) \, \langle 0 | Q_7 | 0 \rangle_\chi (\mu_R) 
\nonumber \\ +
{\rm Im} C_8(\mu_R) \, \langle 0 | Q_8 | 0 \rangle_\chi (\mu_R) \, . &  
\eea
In the chiral limit and in the NDR scheme
(for the HV scheme expressions see \cite{BGP01}),
the matrix elements above  are
\bea 
\langle 0 | Q_7 | 0 \rangle^{\rm NDR}_\chi (\mu_R)  =
 \frac{3}{32 \pi^2} \left( 1 + \frac{1}{24} \frac{\alpha_S}{\pi} \right)
 {\cal A}_{LR}(\mu_R) \nonumber \\  
+ \frac{1}{48} \frac{\alpha_S}{\pi} {\cal B}_{SP}(\mu_R) 
\eea
\bea 
\label{Q8}
\langle 0 | Q_8 | 0 \rangle^{\rm NDR}_\chi(\mu_R)  =
\left( 1 + \frac{23}{12} \frac{\alpha_S}{\pi} \right){\cal B}_{SP}(\mu_R) 
\nonumber \\ 
+ \frac{3}{32\pi^2} \, \frac{9}{2} \frac{\alpha_S}{\pi} {\cal A}_{LR}(\mu_R) 
\eea
with
\be
\label{LR}
{\cal A}_{LR}(\mu_R) \equiv
\int^{s_0}_0 \, {\rm d}t \, t^2 \, \ln \left(\frac{t}{\mu_R^2} \right) \, 
\frac{1}{\pi} \, {\rm Im} \Pi^T_{LR}(t)
\ee
and 
\bea
\label{SP}
{\cal B}_{SP}(\mu_R) \equiv
3  \langle 0 | \overline q q | 0 \rangle^2_{\overline{MS}}(\mu_R) 
\nonumber \\ 
+ \frac{1}{16\pi^2}  
\int^{\hat s_0}_0 \, {\rm d}t \, t \, \ln \left(\frac{t}{\mu_R^2} \right) \, 
\frac{1}{\pi} \, {\rm Im} \Pi^{(0-3)}_{SS+PP}(t)  \, .
\eea
$s_0$ and $\hat s_0$ are the onsets of local QCD duality.
These results are exact in the chiral limit.
The two-point function $\Pi^T_{LR}(Q^2)$ is the transverse part of
$\langle 0| (V^\mu-A^\mu)^{(3)}(x)(V^\nu+A^\nu)^{(3)}(0) | 0 \rangle$
in momentum space, and  
the two-point function $\Pi^{(0-3)}_{SS+PP}(Q^2)$ is the singlet minus
triplet combination of $ \langle 0 | (S+iP)(x) (S-iP)(0) | 0 \rangle$
in momentum space,
both in the chiral limit. See \cite{BGP01} for their explicit expressions.
In this reference, we show
 how scheme and scale dependences of the Wilson coefficients
match analytically the ones of the matrix elements.
This was done there explicitly in the NDR and HV schemes at $O(\alpha_S^2)$.

The first term in (\ref{SP}) comes from a disconnected diagram
and is order $N_c^2$. The second term comes from connected
diagrams and is order $N_c^0$. We also know that 
$\Pi^{(0-3)}_{SS+PP}(Q^2)$ has an OPE which starts at $1/Q^4$ modulated
by $O(\alpha_S^2)$ coefficients times $\langle 0 |Q_7|0\rangle$ and 
$\langle 0 | Q_8 | 0 \rangle$. As a consequence, it fulfills exactly 
a first Weinberg sum rule-like \cite{WEIN75} and very approximately
a second  one \cite{BGP01}.  
Apart of fulfilling the two Weinberg sum rules-like, 
notice that the kernel of the sum rule (\ref{SP}) we are interested in 
has  zeroes at t=0 and at $t=\mu^2$ which we chose to be $\mu^2=\hat s_0$.
One expects therefore that keeping just the first pole
in each channel is a good approximation for estimating
the leading OPE behavior as happens  in 
$\pi^+-\pi^0$ electromagnetic mass difference.

In \cite{BGP01} we discussed this sum rule
using  the known pion and the $\eta_1$ poles and including 
the first pion prime as
a narrow width.  The $\pi'$ contribution
can be improved using  a Breit-Wigner shape
and the results do not change much due to the phenomenologically
small coupling of the $\pi'$.
The scalar counterpart is more delicate and more work is needed.
But as a first estimate, we used  two hadronic models \cite{MOU01,KLM97}
that fulfill the QCD short-distance constraints and 
produce values for $L_4$, $L_6$ and $L_8$ compatible with
phenomenology. The scalar form factor in \cite{MOU01} is obtained
from data and dispersion relations up to 1 GeV and Breit-Wigner
shapes above.
The result of using these models
agreed  with  the results of naive narrow widths for the lowest
scalar resonances. These were constructed to 
fulfill the short-distance QCD constraints
and also produced reasonable values for $L_6$ and $L_8$. 
Here, we have also tried  Breit-Wigner shapes for the scalar
mesons instead of narrow widths 
and again find results in the same ball park.
Now, we also have used the scalar form factors obtained in  \cite{MO01}
where  the lowest scalar triplet and singlet resonances 
are generated dynamically for energies up
to 1 GeV and Breit-Wigner shapes above and we get similar  results.

In all the estimates, we  got negative 
corrections to the first term in (\ref{SP}) 
in the region between $-$10\% to $-$30\%.
Though the scalar-pseudoscalar  sum rule
(\ref{SP}) cannot obviously 
be used at a quantitative level at present, 
the results above  indicate that is difficult to have 
corrections larger than $\pm$30\% 
to the first term in (\ref{SP}).

The chiral limit OPE of $\Pi^T_{LR}(Q^2)$ starts  at $1/Q^6$ 
modulated by 
known coefficients times  $\langle 0 |Q_7|0\rangle$ and 
$\langle 0 | Q_8 | 0 \rangle$. Using this, one arrives at
\cite{DG00,KPdR01,CDGM01,BGP01} 
\bea
\label{M2}
M_2 \equiv 
\int^{s_0}_0 \, {\rm d}t \, t^2 \, \frac{1}{\pi} \, {\rm Im} \Pi^T_{LR}(t)
\simeq \nonumber \\ 
- \frac{4\pi}{3} \alpha_S(s_0) \left(1+\frac{25}{8} \frac{\alpha_S(s_0)}{\pi} 
\right) \, \langle 0 | Q_8 | 0 \rangle_\chi^{\rm NDR} (s_0) \, .
\eea
This is the relation that we use  for
 the quantitative  analysis of $\langle 0 | Q_8 | 0 \rangle$.

This sum rule and the one in (\ref{LR}) can be calculated
using the excellent ALEPH \cite{ALEPH} and  OPAL \cite{OPAL}
tau data. For the details of how we use the tau data we refer to
\cite{BGP01}. Only to mention here that
we have generated around 100,000 tau data distributions according to their
covariance matrix and assigned  to each one  a single value of 
$s_0$. This value is the highest one allowed by data 
where the second Weinberg sum rule (WSR) is fulfilled.
Both sum rules (\ref{LR}) and (\ref{M2}) have $t^2$ as kernel.
This makes the region above 1.5 GeV$^2$ of increasing  importance
with respect to the low energy region.
We have chosen the second WSR since it differs from the one
we are interested  in just by one power less in the kernel.
In addition, the sum rule in (\ref{LR}) has a zero at $t=\mu^2=s_0$. 
We also used the highest value available by the data
-which is always between 2 GeV$^2$ and 3 GeV$^2$-  since one also expects
QCD-Hadron duality  to work better  there.

Combining the ALEPH and OPAL  results, we obtain 
\be
\label{FESR}
M_2=-[1.9\pm1.0] \cdot 10^{-3} \, {\rm GeV}^6
\ee
and always $M_3>0$ 
in agreement with the recent analysis in \cite{Atalks}.

If we use Im$\Pi^{(0-3)}_{SS+PP}(t)=0$ in (\ref{SP})
and  $\langle 0 | \overline q q | 0 \rangle_{\overline{MS}}$(2 GeV) 
= $- (0.018\pm 0.004)$ GeV$^3$  from \cite{BPdR95}, 
we get
\be
\label{SS0}
M_2=-[2.0\pm0.9] \cdot 10^{-3} \, {\rm GeV}^6 \, .
\ee
This result is very compatible with the FESR
analysis result we got in (\ref{FESR}).

For our final result in the NDR scheme,
using $\alpha_S$(2 GeV) = 0.32 (see more details in \cite{BGP01}),
we quote
\begin{eqnarray}
\langle 0 | Q_8 | 0 \rangle^{\rm NDR}_\chi(2 {\rm GeV}) =
\nonumber \\ 
(1.20\pm 0.60 \pm 0.15) \cdot 10^{-3} \, {\rm GeV}^6 =
 \nonumber \\
(1.2 \pm 0.7) \cdot 10^{-3} \, {\rm GeV}^6 \, . 
\end{eqnarray}
Where the first error is purely experimentally 
and takes into account both ALEPH and OPAL results 
as well as a possible variation of the
local duality onset $s_0$ and the second one
is from the unknown $O(\alpha_S^3)$ terms
in (\ref{M2}) assuming a geometrical series.

The sum rule in (\ref{LR}) is much better behaved and with smaller error bars
due to to the zero at $t=\mu^2=s_0$. Using again the same strategy we explained
above for the analysis of sum rule (\ref{M2}), we get 
\be
{\cal A}_{LR}(2 {\rm GeV}) = (4.35 \pm 0.50) \cdot 10^{-3}
\, {\rm GeV}^6
\ee
combining ALEPH and OPAL results.
See \cite{BGP01} for further   details.

Comparison with other recent determinations
is made in Table \ref{results}.
For the results in the cases $B_7^\chi(2 {\rm GeV})=
B_8^\chi(2{\rm GeV})=1$
 and Im$\Pi^{(0-3)}_{SS+PP}(t)=0$
 we used $\langle 0 | \overline q q | 0 \rangle_{\overline{MS}}$
(2 GeV) = $- (0.018 \pm 0.004)$ GeV$^3$ from \cite{BPdR95}, which is 
 in agreement with the most recent sum rule determinations of 
this condensate and of light quark masses -see \cite{JOP02} for instance- 
and the lattice  light quark masses world average in \cite{lattice}.

 Within the present accuracy of $\langle 0 | \overline q q | 0 \rangle$, 
the disconnected contribution to (\ref{SP}) 
-third line in Table \ref{results}- 
is  perfectly compatible with  our
full result -fourth line in Table \ref{results}-
as well as the results from \cite{NAR01,Atalks,Roma2,CPPACS,RBC}
, so that we cannot conclude a  large deviation
from the large $N_c$ result within the present accuracy. Notice that we include
in this result -third line in Table \ref{results}- 
$O(\alpha_S)$ corrections that 
are indeed leading order in $1/N_c$ 
(see (\ref{Q8}) and \cite{BGP01})  which are usually
disregarded in the factorization approaches, this makes the 
chiral limit $B_8^\chi(2 {\rm GeV})$ 
parameter larger than one by around 20\% to 30\%.
\begin{table*}[htb]
\caption{Comparison of NDR results 
for $\mu_R=$ 2 GeV and $n_f=3$ flavors in units of 10$^{-3}$
GeV$^6$.\label{results}}
\vspace{0.4cm}
\renewcommand{\tabcolsep}{1.pc} 
\renewcommand{\arraystretch}{1.15} 
\begin{center}
\begin{tabular}{@{}lll}
 Method & 
$-60 \, \langle 0 | Q_7 | 0 \rangle_\chi$  &
$\langle 0 | Q_8 | 0 \rangle_\chi$ 
\\ \hline \hline 
& & \\
$B_7^\chi$(2 GeV)=
$B_8^\chi$(2 GeV)=1 & $3.2 \pm 1.3$ &  $1.0 \pm 0.4 $ 
\\ 
This Work, \cite{BGP01} 
Im$\Pi_{SS+PP}^{(0-3)}=0$ & $2.4\pm 0.3$  & $1.2 \pm 0.5$    
\\ 
This Work,  \cite{BGP01}
Data \& Duality FESR  &  $2.4\pm 0.3$  & $1.2 \pm 0.7$ 
\\ 
Cirigliano et al., \cite{Atalks}
Data \& Fitted FESR   & $2.1 \pm 0.3$ &  $1.5 \pm 0.3$  
\\ 
Cirigliano et al., \cite{Atalks}
Weighted Data \& Fitted FESR   & $2.2 \pm 0.3$ &  $1.6 \pm 0.4$  
\\ 
Cirigliano et al., \cite{CDGM01}
Weighted Data   & $1.6 \pm 1.0$ & $2.1\pm 0.6$  
\\ 
Knecht et al., \cite{KPdR01}
$N_c \to \infty$, MHA    & $1.1 \pm 0.3$ &  $2.3 \pm 0.7$  
\\  
Narison, \cite{NAR01}
Data \& Tau-like FESR  & $2.1\pm 0.6$ &  $1.4\pm 0.4$ 
\\ 
Donini et al., \cite{Roma2}
Lattice (Wilson)  & $1.5 \pm 0.4$ &  $0.7\pm 0.2$   
\\ 
 CP-PACS Coll., \cite{CPPACS}
Lattice (Chiral) & $2.4 \pm 0.3$  
(stat.)& $1.0\pm0.2$ (stat.) 
\\ 
 RBC Coll., \cite{RBC}
Lattice (Chiral)  & $2.7 \pm 0.3$ (stat.) 
&  $1.1\pm 0.2$ (stat.) 
 \\ 
\end{tabular}
\end{center}
\end{table*}

We thank Bachir Moussallam and Jos\'e Oller for providing us
with scalar form factor programs.
J.P. thanks Vincenzo Cirigliano for helpful discussions. This work 
is supported in part 
by the Swedish Research Council, by the European Union TMR
network, Contract No. HPRN-CT-2002-00311  (EURIDICE), by MCYT (Spain),
Grant No. FPA 2000-1558 and by Junta de Andaluc\'{\i}a, Grant No. FQM-101.
E.G. is indebted to MECD (Spain) for a FPU fellowship.

\end{document}